\documentclass[doublecol]{epl2} 
\usepackage{amsmath}
\usepackage{bbm}
\usepackage[normalem]{ulem}
\usepackage{graphicx}
\usepackage{stfloats}   
\usepackage{perpage} 
\MakePerPage{footnote} 
\def\be{\begin{eqnarray}}
\def\ee{\end{eqnarray}}

\usepackage{xspace}

\def\Al{A}

\definecolor{DarkGreen}{rgb}{0.2,0.8,0.2}

\title{\vspace*{-12mm}Coupling function from bath density of states}
\shorttitle{Coupling Function From Bath Density Of States} 

\author{\vspace*{-05mm}S. Nemati\inst{1} \and C. Henkel\inst{1} \and J. Anders\inst{1,2}}
\shortauthor{S. Nemati \etal}

\institute{                    
  \inst{1} University of Potsdam, Institut f\"ur Physik und Astronomie, 14476 Potsdam, Germany.\\
  \inst{2} Department of Physics and Astronomy, University of Exeter, Stocker Road, Exeter EX4 4QL, UK.
}

\abstract{
Modelling of an open quantum system requires knowledge of parameters that specify how it couples to its environment.
However, beyond relaxation rates, realistic parameters for specific environments and materials are rarely known. 
Here we present a method of inferring the coupling between a generic system and its bosonic (e.g., phononic) environment from the experimentally measurable density of states (DOS). With it we confirm that the DOS of the well-known Debye model for three-dimensional solids is physically equivalent to choosing an Ohmic bath.
We further match a real phonon DOS to a series of Lorentzian coupling functions, allowing us to determine coupling parameters for gold, yttrium iron garnet (YIG) and iron as examples.
The results illustrate how to obtain material-specific dynamical properties, such as memory kernels. 
The proposed method opens the door to more accurate modelling of relaxation dynamics, for example for phonon-dominated spin damping in magnetic materials. 
}

\begin{document}

\maketitle

\section{Introduction} 
Quantum technologies face many challenges, often arising due to the unavoidable coupling of any system to its environment. The prediction of their dynamics requires open quantum system methods that include such coupling effects, for example the Caldeira-Leggett model \cite{Breuer02} and the spin-boson model \cite{Weiss12}. These methods are successfully employed in many physical contexts, e.g., quantum optics \cite{Krinner18, Maniscalco04, Stewart17}, 
condensed matter \cite{Ronzani18, Semin20, Deffner13, Wilhelm04, Hanson08, Luschen17}, 
quantum computation \cite{Verstraete09, Kliesch11, Thorwart01}, 
nuclear physics \cite{Brambilla21}
and quantum chemistry \cite{Teh19}. 
For instance, modelling circuit quantum electrodynamics with the spin-boson model shows that the heat transport of a superconducting qubit within a hybrid environment 
changes significantly, depending on the qubit-resonator and resonator-reservoir couplings \cite{Ronzani18}.

In the mathematical treatment of an open quantum system, a coupling function $\tens{\cal C}_{\omega}$ is typically introduced that describes how strongly the system interacts with bath degrees of freedom (DoF). 
Its functional form determines the temporal memory of the bath and whether the noise is coloured or not \cite{Breuer02,Weiss12,Anders20}, 
critically affecting the system dynamics \cite{Deffner13, Liu16, Zou20}.
A large body of theoretical results exist for various toy models that make specific assumptions on the coupling function $\tens{\cal C}_{\omega}$ \cite{Weiss12, Breuer02, Shabani05}. 
However, a major drawback is a somewhat lacking connection to system- or material-specific characteristics to which these methods could be applied: for a given DoF, in a given material, which coupling function $\tens{\cal C}_{\omega}$ should one choose to model its dynamics?

An alternative approach is taken in the condensed matter literature, where open quantum systems are usually characterized by the density of states (DOS) of their environment \cite{Chen05}. Measurement of, for example, the phonon DOS is well-established using different inelastic scattering techniques \cite{Ament11,Bayle14}. Modes in the environment typically couple to the system with a wave vector-dependent strength $g_{\vect{k}}$ \cite{Nazir16, Calarco03, Weiss12}, which in many cases can be captured by a frequency-dependent $g_{\omega}$.

In this paper, we present a useful relation that translates the coupling function $\tens{\cal C}_{\omega}$ of an open quantum system into an experimentally measurable DOS $D_{\omega}$, and vice versa.
While a similar relation has previously been reported for one-dimensional quantum spin impurities \cite{Bulla97, Bulla05}, the relation obtained here is valid for a generic system coupled to a bosonic bath, capturing dimensionality and anisotropy. 
It paves the way to parametrizing realistic coupling functions for a range of applications, for example, for spins in a magnetic material that experience damping through the coupling to the crystal lattice \cite{Anders20, Reppert20}
or for nitrogen vacancy centers, a solid-state analogue of trapped atoms, whose coherence lifetime in optical transitions is also limited by interaction with phonons \cite{Bar-Gill12,Fuchs10}.
The link is explicitly established for a generic quantum system that couples locally to a bosonic environment. Extensions to other environments, such as fermionic environments, will be possible using similar arguments. 

The paper is organised as follows: 
we first introduce the two approaches involving $D_{\omega}$ and $\tens{\cal C}_{\omega}$, respectively. 
Setting up the dynamics of the environment, we evaluate its memory kernel and establish the link between $D_{\omega}$ and $\tens{\cal C}_{\omega}$, allowing for general $g_{\omega}$.
In the second part of the paper, we choose a flat $g$ for simplicity, and illustrate the application of the relation with a few examples. We demonstrate that the widely
used Debye approximation is equivalent to the well-known Ohmic coupling function. 
While this approximation suffices at low frequencies, experimental DOS show peaks at higher frequencies, leading to non-trivial dissipation regimes. 
We parametrize two measured phonon DOS, those of gold and iron (see Supplementary Material (SM)), and one theoretically computed phonon DOS of yttrium iron garnet (YIG) and extract key parameters for the corresponding coupling functions $\tens{\cal C}_{\omega}$.
These give direct insight into the impact of memory for any phonon-damped dynamics in these materials. 

\begin{figure}
\includegraphics[width=0.47\textwidth]{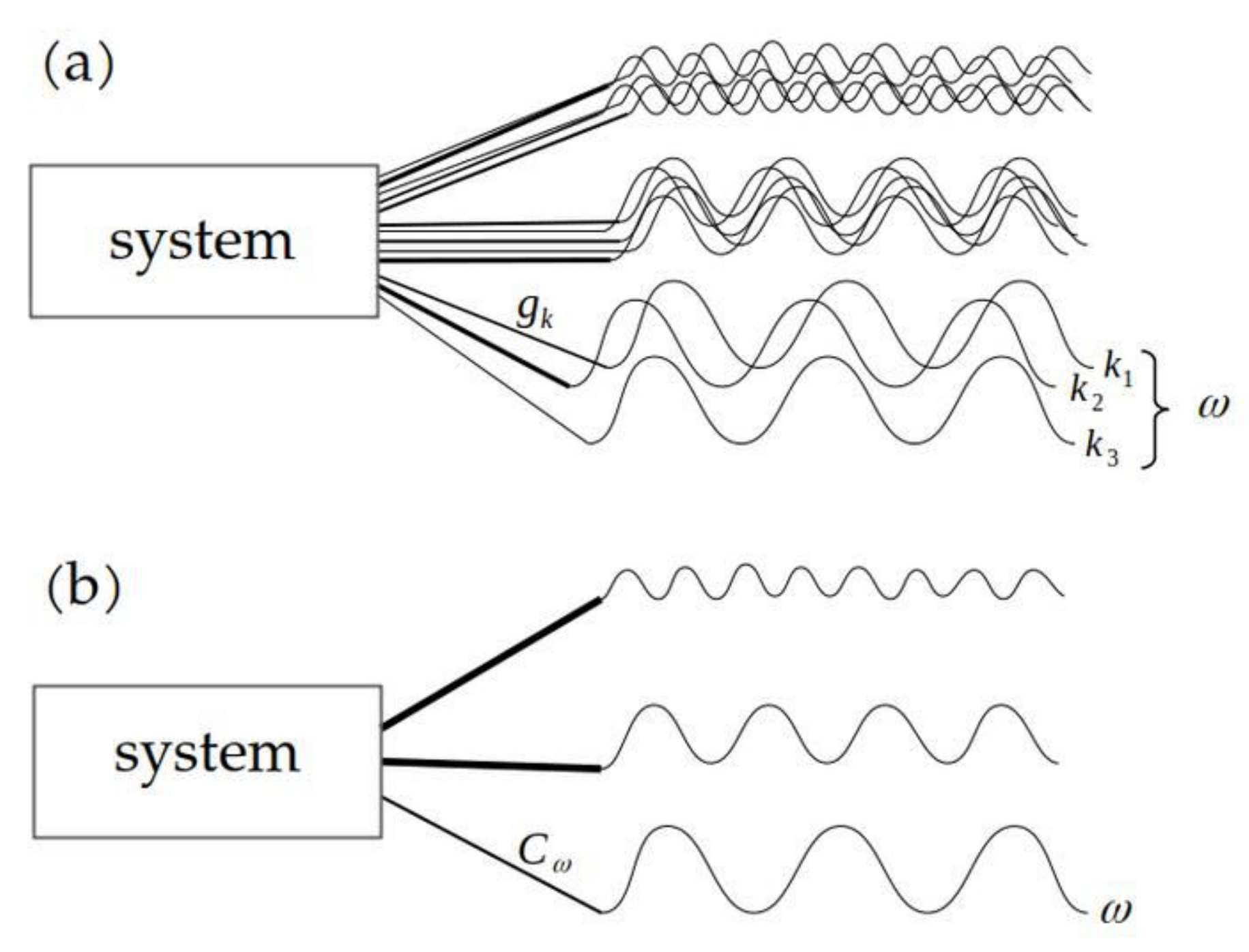}
\caption{Schematic picture of two equivalent approaches to modelling the open quantum systems. (a) Wave vector approach: Each bath frequency $\omega$ includes several wave vectors $\{\vect{k}\}$
where each bath wave vector $\vect{k}$ couples to the system with strength $g_{\vect{k}}$.
(b) Frequency approach: Every bath frequency $\omega$ couples to the system with a strength given by $\tens\cal C_{\omega}$.}
\label{fig1}
\end{figure}

\section{Two approaches}\label{two-appraoches}
The Hamiltonian of a quantum system in contact with a 
bath is
\begin{eqnarray}
\hat{{\cal H}}_{tot} = \hat{{\cal H}}_S+ \hat{{\cal H}}_B +\hat{{\cal H}}_{SB}\,,
\label{eq:H-tot}
\end{eqnarray}
where the bath Hamiltonian $\hat{{\cal H}}_B$ and the system Hamiltonian $\hat{{\cal H}}_S$ may contain the internal interactions among their own components. 
The system-bath interaction is assumed to be of product form,
\begin{eqnarray}
\hat{{\cal H}}_{SB} = - \hat{\vect{S}} \cdot \hat{\vect{B}}\,,
\label{eq:system-bath-H}
\end{eqnarray}
where $\hat{\vect{S}}$ is a (Hermitian) system operator and $\hat{\vect{B}}$ is a bath operator, each with $d_s$ components. The form of the bath Hamiltonian $\hat{{\cal H}}_{B}$ and of the bath operator $\hat{\vect{B}}$ depends on the context. We consider here a bosonic bath, i.e. an infinite set of harmonic oscillators.
In the literature, one can broadly distinguish two representations of the bath, working either in wave vector (WV) or frequency (F) \ space, as illustrated in Fig.\,\ref{fig1}. 

The wave vector approach is common in condensed matter physics \cite{Chen05, Weiss12} where the bath Hamiltonian is expressed as a sum over all possible 
modes $\vect{k}$
\begin{eqnarray}
\hat{\cal H}_B^{WV} &=& \sum_{\vect{k}} \hbar \omega_{\vect{k}}\left(\hat{b}_{\vect{k}}^\dag \hat{b}_{\vect{k}}+\frac{1}{2}\right)\,.
\label{eq:bath-H}
\end{eqnarray}
Here $\omega = \omega_{\vect{ k}}$ gives the dispersion relation of a normal mode with wave vector $\vect{k}$
and $\hat{b}_{\vect{k}}$ ($\hat{b}_{\vect{k}}^{\dag}$) are bosonic annihilation (creation) operators of a mode excitation with commutation relations $[ \hat{b}_{\vect{k}},\hat{b}_{\vect{k}'}^{\dag}] = \delta_{\vect{k} \vect{k}'}$.
Usually one considers a three-dimensional ($3$D) structure with wave vectors $\vect{k} = (k_x, k_y, k_z)$. For example, in a cubic $3$D lattice with number of lattice sites $N$, lattice constant $a$ and volume $V = Na^3$, each component of $\vect{k}$ runs through the range $\left(-\frac{\sqrt[3]{N}-1}{2}, \ldots, 0, \ldots, \frac{\sqrt[3]{N}-1}{2} \right) \, \frac{2 \pi}{\sqrt[3]{N} a}$. 
For large $N$ and $V$, and 
for any function $f(\omega_{\vect{k}})$ that only depends on the frequency $\omega_{\vect{k}}$, one can approximate sums over the wave vectors as
\begin{align}
\frac{1}{V}
\sum_{ \vect{k}} f(\omega_{\vect{k}}) 
\cong  \, 
\int \! \frac{\upd^3k}{(2\pi)^3} \, f(\omega_{\vect{ k}}) 
=: \int \!\upd\omega \,
D_{\omega} \, f(\omega)\,.
\label{eq:l-omega}
\end{align}
This equation defines $D_{\omega}$ as the DOS per unit volume of bath modes at frequency $\omega$ \cite{Chen05}.

For bosonic baths, we choose the standard interaction \cite{Weiss12} where the bath operator $\hat{\vect{B}}$ is linear in the bosonic mode operators (single phonon processes), 
\begin{eqnarray}
\hat{\vect{B}}^{WV} = \frac{1}{\sqrt{V}}\sum_{\vect{k}} \vect{ \xi}_{\vect{k}} \, \hat{b}_{\vect{ k}} 
+ \text{h.c.}\,,
\label{eq:phonon-field-2}
\end{eqnarray}
where $\vect{ \xi}_{\vect{k}} = \vect{ \epsilon}_{\vect{ k}} \, \left( \hbar g_{\vect{k}}^2/ (2 \omega_{\vect{k}}) \right)^{1/2}$ with $\vect{ \epsilon}_{\vect{ k}}$ a $d_s$-dimensional unit polarisation vector \cite{Breuer02} and $g_{\vect{k}}$ the wave vector-dependent coupling, see Fig.~\ref{fig1}.
Eq.\,\eqref{eq:system-bath-H} may be generalized to the situation that several system components $\hat{\vect{S}}_m$ are located at different positions $\vect{R}_m$, and sum over interaction terms, i.e. $\hat{{\cal H}}_{SB} = -\sum_{m}\hat{\vect{S}}_m \cdot \hat{\vect{B}}( \vect{R}_m )$. The field operators would then be $\vect{R}$-dependent, i.e. $\hat{\vect{B}}^{WV}(\vect{R}) = \frac{1}{\sqrt{V}}\sum_{\vect{k}} \vect{ \xi}_{\vect{k}} \, \hat{b}_{\vect{ k}} \, {\rm e}^{ {\rm i} \vect{k} \cdot \vect{R} } + \text{h.c.}$.
For simplicity, we will concentrate in the following on just one system site and drop summation over $m$ again.

Another approach to setting up the bath Hamiltonian $\hat{{\cal H}}_B$ and the interaction $\hat{{\cal H}}_{SB}$ is based on a frequency expansion  often employed in the open quantum systems literature \cite{Weiss12, Breuer02}.
In contrast to Eq.\,\eqref{eq:bath-H}, 
here $\hat{{\cal H}}_B$ is written directly as a sum or integral over frequencies,
\begin{eqnarray}
\hat{{\cal H}}_B^{F} = \frac{1}{2}\int_{0}^{\infty} \!\!\!\!\upd\omega \left( \hat{\vect{ P }}_{\omega}^2 + \omega^2\hat{\vect{ X}}_{\omega}^2\right), 
\label{eq:HB-omega}
\end{eqnarray}
where $\hat{\vect{ P }}_{\omega}$ and $\hat{\vect{X}}_{\omega}$ are $3$D
[in general, $d$-dimensional ($d$D)] momentum and position operators, respectively, for the bath oscillator with frequency $\omega$. Their components obey $[\hat{X}_{\omega,j}, \hat{P}_{\omega',l}] = {\rm i}\hbar \,\delta_{jl}\,\delta(\omega - \omega')$.  
In this approach, the bath operator in Eq.\,\eqref{eq:system-bath-H} is often chosen as \cite{ Anders20}
\begin{eqnarray}
\hat{\vect{B}}^{F} = \int_{0}^{\infty} \!\!\!\!\upd\omega \,\tens{\cal C}_{\omega}  \hat{\vect{ X}}_{\omega}\,,
\label{eq:B-omega}
\end{eqnarray}
where the coupling function $\tens{\cal C}_{\omega}$ (in general a $d_s\times d$ tensor) is weighting the system-bath coupling at frequency $\omega$. 
{The system operators couple isotropically to the bath if $\tens{\cal C}_{\omega}\tens{\cal C}_{\omega}^{T} =  \mathbbm{1}_{d_s}\,C_{\omega}^2$. The scalar coupling function $C_{\omega}$}
is related to the bath spectral density $J_{\omega}$, which alternatively quantifies the effect of the environment on the system as $J_{\omega} \propto C^2_{\omega}/\omega$ \cite{Breuer02,Weiss12}. The bath dynamics can be categorised \cite{Weiss12} based on 
the low-$\omega$ exponent of the spectral density, 
$J_{\omega}\propto \omega^s$, into three different classes, called Ohmic ($s = 1$), sub-Ohmic ($s < 1$), and super-Ohmic ($s > 1$).

The difference between wave vector approach and frequency approach is that at a fixed frequency $\omega$, 
there is in 
Eq.\,\eqref{eq:B-omega} just one bath operator $\hat{\vect{X}}_\omega$ that couples to the system, 
while according to Eq.\,\eqref{eq:phonon-field-2}, 
the interaction is distributed over several wave vector modes $\vect{k}$ with weighting factors $\vect{ \xi}_{\vect{ k}}$, their number being set by the DOS $D_{\omega}$ (see Fig.\,\ref{fig1}). 

We now want to address the question of the connection between the DOS $D_{\omega}$ and the coupling function $\tens{\cal C}_{\omega}$. To achieve this, we consider one relevant quantity in both approaches and equate the corresponding formulas. 
In the following, we choose the memory kernel $\tens{\cal K}$ which encodes the response of the bath to the system operator $\hat{\vect{S}}$.
Note that the choice of $\hat{\vect{B}}$ in Eq.\,\eqref{eq:phonon-field-2} restricts the discussion to the linear response of the bath, as is reasonable for a bath that is thermodynamically large \cite{Breuer02, Weiss12}. 

\section{Memory kernel in both approaches} 
To find an explicit relation in the wave vector approach for the dynamics of the bath operator $\hat{{\vect{B}}}^{WV}$ in Eq.\,\eqref{eq:phonon-field-2}, the starting point is the equation of motion for $\hat{b}_{\vect{ k}}$,
\begin{eqnarray}
\frac{ d \hat{b}_{\vect{ k}} }{ dt } = - {\rm i} \omega_{\vect{k}} \hat{b}_{\vect{ k}} 
+ \frac{{\rm i} }{\hbar \sqrt{V}} \vect{ \xi}_{\vect{k}}^{\dag} \cdot \hat{{\vect{S}}}
\, ,
\label{eq:equation-of-motion}
\end{eqnarray}
whose retarded solution contains two terms
\begin{eqnarray}
    \hat{b}_{\vect{k}}(t) & = & \hat{b}_{\vect{k}}(0) \, {\rm e}^{- {\rm i} \omega_{\vect{k}} t} 
\\
    && {} + \frac{{\rm i} }{\hbar \sqrt{V}} 
    \vect{ \xi}_{\vect{k}}^{\dag} \cdot
    \int_{0}^{t}\!\!\upd t'\,  \hat{{\vect{S}}}(t') \, {\rm e}^{- {\rm i} \omega_{\vect{k}} (t-t')}\,. \nonumber
\label{eq:solution-equation-of-motion}
\end{eqnarray}
Therefore, the time evolution of the bath operator can be written as
$\hat{{\vect{B}}}^{WV}(t) = \hat{{\vect{B}}}_{\rm induced}^{WV}(t) + \hat{{\vect{B}}}_{ \rm  response}^{WV}(t)$. 
The first term represents the internally evolving bath which is given by $\hat{{\vect{B}}}_{\rm  induced}^{WV}(t)
= \frac{1}{\sqrt{V}}\sum_{\vect{k}}\hat{b}_{\vect{k}}(0) e^{-i\omega t}\vect{ \xi}_{\vect{k}}
+ \text{h.c.}$, 
while ${\hat{\vect{B}}}_{\rm response}^{WV}(t)$ contains information about the system's past trajectory,
\begin{eqnarray}
\hat{{\vect{B}}}_{\rm  response}^{WV}(t)
= 
\int_{0}^{\infty}\!\!\!\!\upd t'\,
\tens{\cal K}^{WV}(t-t') \,\hat{{\vect{S}}}(t')\,,
\label{eq:response-field}
\end{eqnarray}
where $\tens{\cal K}^{WV}(t-t')$ is the memory kernel (a tensor),
\begin{eqnarray}
\tens{\cal K}^{WV}(t-t') = 
\frac{\Theta(t-t')}{ V}\sum_{\vect{k}} g_{\vect{k}}^2
\vect{ \epsilon}_{\vect{k}} \vect{ \epsilon}_{\vect{k}}^{\dag}
\frac{\sin \omega_{\vect{k}}(t-t')}{\omega_{\vect{k}}}\,.
\label{eq:kernel-1}
\end{eqnarray}
Here, the $\vect{ \xi}_{\vect{k}}$ have been expressed by the unit polarisation vectors $\vect{ \epsilon}_{\vect{k}}$ [see after Eq.\,\eqref{eq:phonon-field-2}] and $\Theta(t-t')$ is the Heaviside function, which ensures that the bath responds only to the past state of the system, i.e. $t' < t$. 

For large volume $V$, the summation over $\vect{k}$ in Eq.\,\eqref{eq:kernel-1} can be transformed into a frequency integration as in  Eq.\,\eqref{eq:l-omega}. 
The projection on polarization vectors, averaged over an isofrequency surface $\Omega$, is taken into account by a  ($d_s \times d_s$) positive Hermitian matrix $g_{\omega}^2 \tens{{\cal M}}_{\omega} = (\Omega)^{-1}\int \upd \Omega \, g_{\vect{k}}^2 \; \vect{ \epsilon}_{\vect{k}} \vect{ \epsilon}^{\dag}_{\vect{k}}$, where the matrix $\tens{{\cal M}}_{\omega}$ is normalized to unit trace and $g_{\omega}^2$ is a scalar. With these notations, the memory tensor in the wave vector approach is 
\begin{eqnarray}
\tens{\cal K}^{WV}(t-t') = 
\Theta(t-t')
\!\!\int_{0}^{\infty}\!\!\!\!\upd\omega\,
g_{\omega}^2\tens{{\cal M}}_{\omega}
D_\omega 
\frac{\sin \omega(t-t')}{\omega}
\,.
\label{eq:kernel1}
\end{eqnarray}

Turning now to the frequency approach, the dynamics of the bath operator $\hat{\vect{X}}_{\omega}$ in Eq.\,\eqref{eq:B-omega} follows a driven oscillator equation 
\begin{eqnarray}
\frac{d^2\hat{\vect{X}}_{\omega}}{dt^2}+\omega^2\hat{\vect{X}}_{\omega} =  \tens{\cal C}_{\omega}^{T}\, \hat{\vect{S}}
\, .
\label{eq:equation-of-motion-X}
\end{eqnarray}
Its exact solution is 
\begin{eqnarray}
\hat{\vect{X}}_{\omega}(t)&=& \hat{\vect{X}}_{\omega}(0)\cos{\omega t} + \hat{\vect{ P }}_{\omega}(0)\sin{\omega t}\nonumber\\
&&{} +  \int_{-\infty}^{\infty}\!\!\!\!\upd t'G_{\omega}(t-t')\, \tens{\cal C}_{\omega}^{T}\,  \hat{\vect{S}}(t')\,,
\label{eq:X-solution}
\end{eqnarray}
where $G_{\omega}(t-t') = \Theta(t-t') \sin \omega(t-t')/\omega$ 
is the retarded Green's function. Inserting this solution in Eq.\,\eqref{eq:B-omega} leads again to induced and response evolution parts given, respectively, by $\hat{\vect{B}}_{\rm induced}^{F}(t)= \int_{0}^{\infty} \!\!\upd\omega\left(\hat{\vect{X}}_{\omega}(0)\cos{\omega t} + \hat{\vect{ P }}_{\omega}(0)\sin{\omega t}\right)$
and 
\begin{eqnarray}
\hat{\vect{B}}_{\rm  response}^{F}(t)=  \int_{0}^{\infty} \!\!\!\!\upd\omega\int_{0}^{\infty}\!\!\!\!\upd t'\,G_{\omega}(t-t') \,\tens{\cal C}_{\omega}\tens{\cal C}_{\omega}^{T}  \,\hat{\vect{S}}(t')\,.
\label{eq:B-1-omega}
\end{eqnarray}
Comparing with Eq.\,\eqref{eq:response-field} one can identify the memory kernel tensor in the frequency approach as
\begin{eqnarray}
\tens{\cal K}^{F}(t-t') =  \int_{0}^{\infty} \!\!\!\!\upd\omega \; \tens{\cal C}_{\omega}  \tens{\cal C}_{\omega}^{T} \,G_{\omega}(t-t')\,.
\label{eq:kernel2}
\end{eqnarray}

\section{Coupling function $\tens{\cal C}_{\omega} $ versus DOS $D_{\omega}$}
Since Eqs.\,\eqref{eq:kernel1} and \eqref{eq:kernel2} describe the same memory effects, we may set them equal,  leading to
\begin{eqnarray}
 \tens{\cal C}_{\omega}  \tens{\cal C}_{\omega}^{T} 
= 
g_{\omega}^2\,\tens{\cal M}_{\omega} D_{\omega}
\,.
\label{eq:Dw-Cw}
\end{eqnarray}
This relation links the system-bath couplings in the two approaches, i.e. the DOS $D_{\omega}$ is proportional to the Hermitian "square" of the coupling function $\tens{\cal C}_{\omega}$. 
This is the first result of the paper. 

The result\,\eqref{eq:Dw-Cw} may be applied to any quantum system that interacts linearly with a bosonic bath. For instance, magnetic materials in which spins $\hat{\boldsymbol S}$ relax in contact with a phonon reservoir have been studied extensively \cite{Thorwart01, Costi03, Anders20}. The noise-affected occupation of fermionic modes in a double quantum dot \cite{Purkayastha2021}, and the behaviour of an impurity in a Bose-Einstein condensate environment \cite{Lampo17} are other examples.

Note that in Eq.~\eqref{eq:Dw-Cw} the dimension of the system is either smaller or equal to the dimension of the bath, i.e. $d_s \leq d$. A rectangular ($d_s~\times~d$) coupling matrix $\tens{\cal C}_{\omega}$ may model a graphene-on-substrate structure, where the electronic system ($d_s = 2$) is in contact with a $3$D phononic bath \cite{Cusati17}.
An example for equal dimensions is a $3$D spin vector that couples to a $3$D phononic environment \cite{Anders20}.

\section{Specific examples}

In this second part of the paper, we wish to use Eq.\,\eqref{eq:Dw-Cw} to obtain coupling function estimates from experimentally measurable quantities. To do so we have to drop generality and make a number of simplifying assumptions. 

First, we assume isotropic coupling to an isotropic bath, and set $\tens{\cal C}_{\omega} = \mathbbm{1}_{d_s}\,C_{\omega}$ with scalar $C_{\omega}$, and $\tens{\cal M}_{\omega} =  \mathbbm{1}_{d_s}/d_s$.
Second, for simplicity, we assume a frequency-independent $g$ so that the frequency-dependent impact of the coupling is captured by $D_{\omega}$ alone. This is a common approximation for quantum optics systems \cite{Carmichael09, CDG2}, while examples of condensed matter systems where this approximation holds over a range of frequencies are limited \cite{Weiss12}. 
Whenever a non-trivial $g_{\omega}$ is known for a specific context, such as a power law behaviour $\propto \omega^p$, this can be included in Eq.~\eqref{eq:Dw-Cw} separately from the DOS' $\omega$-dependence. To establish such $g_{\omega}$ requires microscopic models for specific physical situations, which make several approximation steps. For example, a derivation of the  electron-phonon interactions in quantum dots \cite{Calarco03} was given in \cite{Nazir16}.

These assumptions reduce Eq.\,\eqref{eq:Dw-Cw} to the scalar equation 
\begin{eqnarray}
C_{\omega}^2 = \frac{g^2}{d_s} D_{\omega}\,,
\label{eq:scalar-Dw-Cw}
\end{eqnarray}
with system dimension $d_s$. 
We will base the following discussion of examples for the coupling functions $C_{\omega}$ on this simpler scalar form.

\section{Debye approximation} 

\begin{figure}[bt]
    \begin{center}
\includegraphics[width=0.4\textwidth]{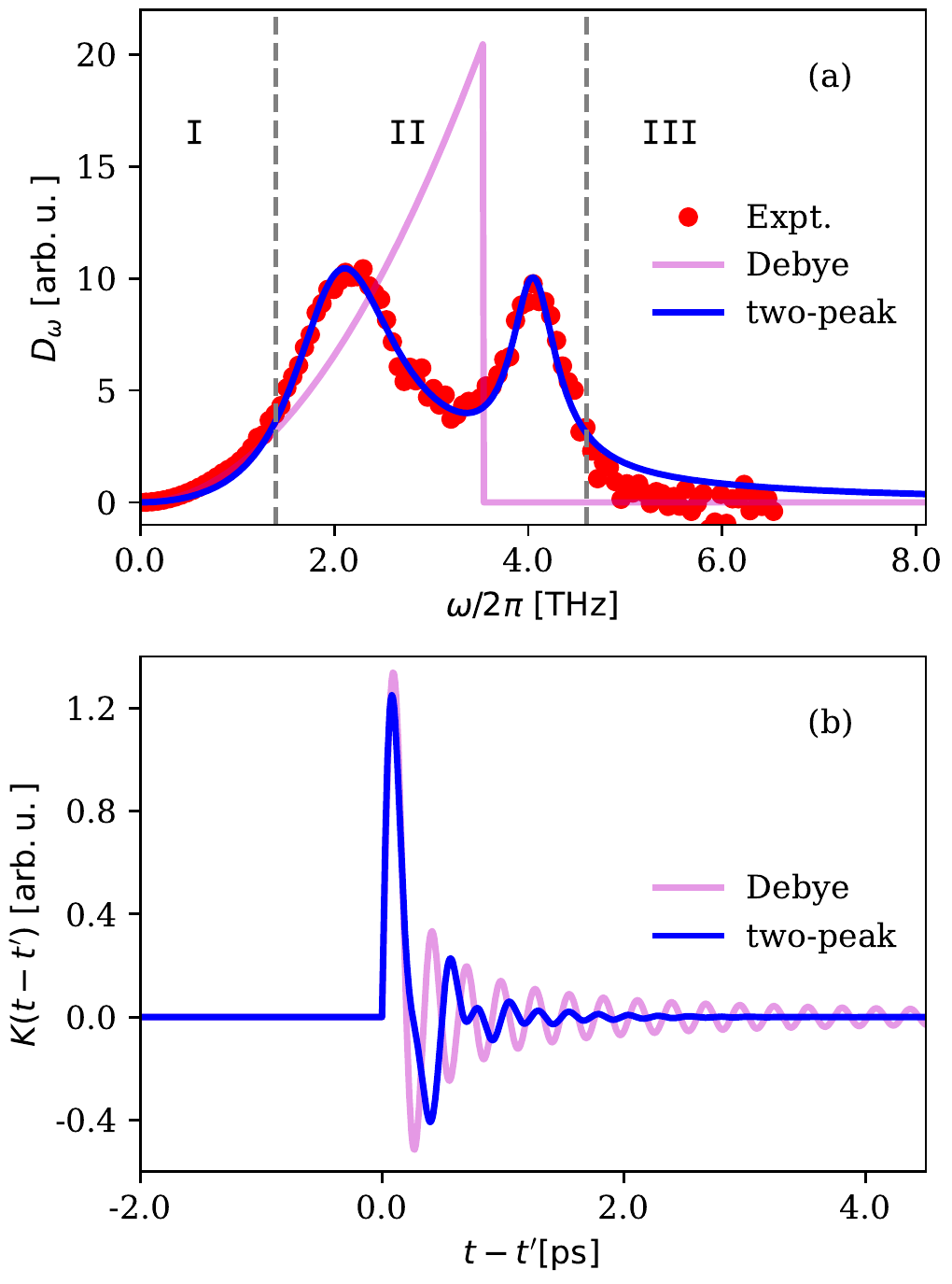}
\caption[]{
(a) Debye DOS (pink solid line, Eq.\,\eqref{eq:Debye-DOS}) and two-peak Lorentzian DOS (blue solid line, Eq.\,\eqref{eq:Dw-Lorentzian-sum}) fitted to a measured phonon DOS for gold (red dots) reported as in Ref.\cite{Munoz13}. The Debye frequency for gold is $\omega_D/2\pi = 3.54\un{THz}$ given in Ref.\cite{Chen05}. Fit specified peak frequencies $\omega_{0,j}$, widths $\Gamma_{j}$ and peak ratios $A_j/A_1$ are given in Table~\ref{tab:fit-to-Au}. The grey dashed lines separate three frequency regimes discussed in the main text. (b) Memory kernels ${\cal K}(t-t')$ corresponding to Debye DOS and two-peak Lorentzian DOS. 
}
\label{fig2}
\end{center}
\end{figure}

In condensed matter physics, the Debye model is used to describe the phonon contribution to a crystal's thermodynamic properties. It assumes an acoustic dispersion, i.e. $\omega = c \vert \vect{k}\vert$ with an averaged sound speed $c$, resulting in $3$D in \cite{Chen05}
\begin{eqnarray}
D_{\omega}^{\rm Deb} = \frac{3\,\omega^2}{2\pi^2 c^3}\, \Theta(\omega_{\rm D} - \omega)\,.
\label{eq:Debye-DOS}
\end{eqnarray}
Here $\omega_D$ is the Debye frequency, i.e. the maximum bath frequency, which in practice is taken to be near the edge of the Brillouin zone.  For example, for gold, see Fig.\,\ref{fig2} (a), the Debye model fits the DOS data reasonably well in frequency region~$I$ up to $\approx 1.4 \un{THz}$.

For the Debye DOS, our relation Eq.\,\eqref{eq:scalar-Dw-Cw} implies the coupling function (setting $d_s = d = 3$)
\begin{eqnarray}
C_{\omega}^{\rm Deb} = \frac{g\,\omega}{\sqrt{2\pi^2 c^3}} \, \Theta(\omega_{\rm D} - \omega)\,.
\label{eq:Debye-Cw}
\end{eqnarray}
The scaling of $C^{\rm Deb}_{\omega}$ implies that the spectral density $J(\omega) \propto C_{\omega}^2/\omega$ is Ohmic, i.e. $J(\omega) \propto \omega$. Hence, the 3D Debye model with constant coupling $g$ in the wave vector approach captures the same relaxation dynamics as an Ohmic bath in the frequency approach. 

Beyond $3$D cubic lattices, $D_{\omega}$ will depend on the dimensionality and lattice symmetry. 
What happens if the lattice is effectively two- or one-dimensional? To answer this, let us imagine a $d$D isotropic lattice with volume $V=Na^d$.
The volume element of such a lattice in $\vect{k}$-space corresponds to $\upd^{d}k = \Omega_d k^{d-1} \upd k$ where $\Omega_d = 2, 2\pi, 4\pi$ 
is the $d$D solid angle for $d = 1, 2, 3$, respectively.

Analogously to the $3$D lattice, using the acoustic dispersion with an averaged sound speed $c$,
one finds the $d$D Debye DOS
\begin{eqnarray}
D_{\omega}^{(d)} = 
\frac{ \Omega_d\, \omega^{d-1} }{ (2 \pi c )^d }
\Theta(\omega_{\rm D} - \omega)\,.
\label{eq:Dw-in-dD}
\end{eqnarray}
Via Eq.\,\eqref{eq:scalar-Dw-Cw} we obtain the power-law
$C_{\omega}\propto \omega^{(d-1)/2}$ for the corresponding coupling functions which implies spectral densities $J(\omega) \propto \omega^{d-2}$.
Thus, isotropic baths in $2$D or $1$D behave in a distinctly sub-Ohmic way.

\section{Inferring coupling functions from DOS data}

Here we wish to go beyond the conceptually useful Debye model, and fully specify the functional form of $C_{\omega}$, given experimentally accessible DOS data that characterise the phononic environment. 

\begin{figure}[t]
 \begin{center}
 \includegraphics[width=0.45\textwidth]{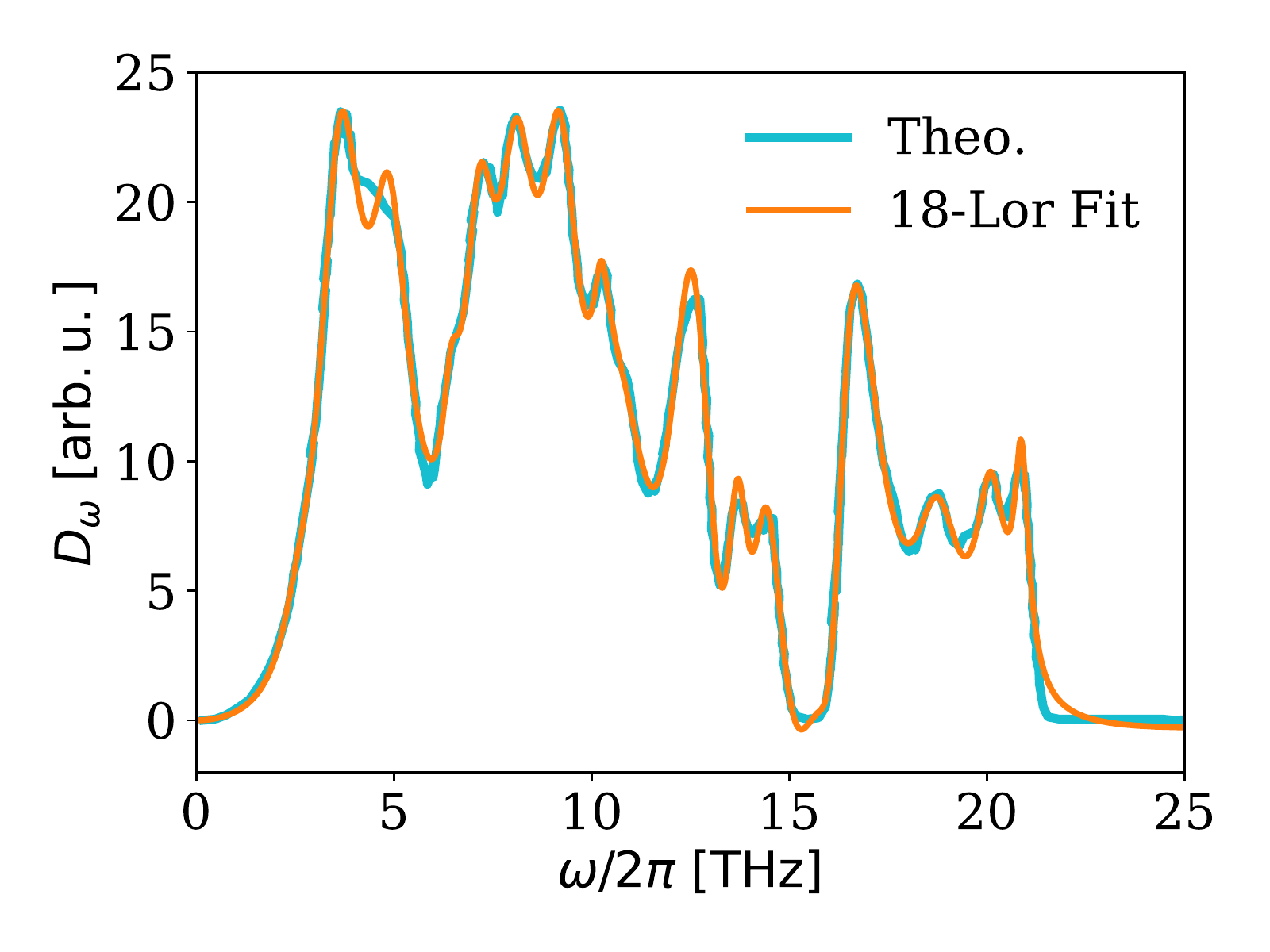}
 \caption[]{ Illustration of eighteen-peak Lorentzian DOS, Eq.\,\eqref{eq:Dw-Lorentzian-sum}, (orange curve) fitted to the theoretically predicted phonon DOS $D_{\omega}$ for YIG (cyan curve) reported in Ref.\cite{Wang20}. The grey dashed line shows a single-peak Lorentzian fit. The fitted
 peak frequencies $\omega_{0,j}$, widths $\Gamma_{j}$ and amplitude ratios $A_j/A_1$ can be found in Table~2 in the SM.}
 \label{fig3}
 \end{center}
 \end{figure}

A generic feature of real materials is a structured DOS, which shows several peaks \cite{Munoz13,Mauger14}. 
Sums of Lorentzian or Gaussian functions are two convenient candidates to approximate such peaky shaped densities \cite{Lemmer18}. Here, we fit experimentally measured DOS for gold \cite{Munoz13} (and iron \cite{Mauger14} in SM) and theoretically computed DOS for YIG \cite{Wang20} to a function consisting of multiple Lorentzians,
\begin{eqnarray}
D_{\omega}^{\rm Lor} = \frac{6\, A_1}{g^2\pi} \sum_{j=1}^{\nu} \frac{A_{j}\Gamma_{j}}{A_1}\frac{\omega^2}{(\omega_{0,j}^2 - \omega^2)^2+\Gamma_{j}^2 \omega^2}\,.
\label{eq:Dw-Lorentzian-sum}
\end{eqnarray}
 The fits, see Figs.\,\ref{fig2} (a), \ref{fig3} and figure in SM, reveal the material specific peak frequencies $\omega_{0,j}$, peak widths $\Gamma_j$ and peak ratios  $A_j/A_1$, see Table\,\ref{tab:fit-to-Au} and tables in SM, while the first peak amplitude $A_1$ remains undetermined. Fixing $A_1$ would require information additional to the DOS, such as the system's relaxation rate due to interaction with the phonon bath. Note that phonon DOS are generally  slightly temperature dependent \cite{Mauger14}. Hence the fit parameters in Eq.\,\eqref{eq:Dw-Lorentzian-sum} will be (usually weak) functions of temperature, a dependence that only matters when a large range of temperatures is considered.

\begin{table}[bhtbp]
    \centering
    \caption{Fit parameters of two-peak Lorentzian matched to the experimentally measured DOS for gold reported in Ref.\cite{Munoz13} (see Fig.~\ref{fig2} (a)).}
    \vspace{2mm}
    \begin{tabular}{c|ccc}
         \hline
         \text{peak}& \text{frequency} & \text{width} & \text{ratio}\\
        $j$ & $\omega_{0,j}/2\pi\ [\un{\!THz}]$ & $\Gamma_{j}/2\pi\ [\un{\!THz}]$ & $A_j/A_1$\\
        \hline
        1 & 2.11 & 1.3 & 1 \\
        2 & 4.05 & 0.56 & 0.15 \\
                \hline
    \end{tabular}
    \label{tab:fit-to-Au}
\end{table}

The peak widths in Eq.\,\eqref{eq:Dw-Lorentzian-sum} determine a characteristic memory time $1/\Gamma_j$. However, beyond this single timescale number, the functional dependence of the memory is fully determined by the kernel  Eq.\,\eqref{eq:kernel1}, which for multi-peak Lorentzians is proportional to
\begin{equation}
\tens{\cal K}^{\rm Lor}(t-t') \propto \sum_{j}^{\nu} A_j e^{-\frac{\Gamma_{j}(t-t')}{2}} \frac{\sin (\omega_{1,j}(t-t'))}{\omega_{1,j}}
\Theta(t-t')\,,
\end{equation}
with $\omega_{1,j} = \sqrt{\omega_{0,j}^2 - \Gamma_j^2/4}$. 
The degree of memory introduced by this kernel into a system's dynamics could be quantified in terms of several non-Markovianity measures, see e.g. \cite{Breuer09, Rivas10, Guarnieri16,Szrtrikacs21}.

For gold, Fig.\,\ref{fig2} (a) shows the phonon DOS measured by Mu\~{n}oz {\it et al.} \cite{Munoz13}, together with our two-peak Lorentzian fit.
The fit gives good agreement in all frequency regimes, with a slightly slower decay in region $III$ than the measured DOS. 
For a system coupled to phonons in gold,
the peak widths (see Table~\ref{tab:fit-to-Au}) immediately imply a characteristic memory time in the picosecond range. 
The relevant kernel is shown (blue) in Fig.\,\ref{fig2} (b) for the two-peak fitted DOS of gold shown in~(a). 
Using the Debye model instead would give a qualitatively different behaviour: the pink curve shows a distinctly slower long-time tail, due to the sharp cutoff at the Debye frequency. 
Note also that without any cutoff, the kernel would be $\tens{\cal K} (t-t') \propto \partial_{t'}\delta (t - t')$, implying no memory \cite{Anders20}. 
In contrast, the Lorentzian fit (blue) provides a quantitatively accurate memory kernel. 

Our approach may provide a more realistic picture of the magnetization dynamics based on actual material data. YIG \cite{Barker20, Barker21} is a typical magnetic insulator in which the relaxation of a spin DoF $\hat{\boldsymbol S}$ is dominated by the coupling to phonons\cite{Sebastian18}, similar to magnetic alloys like Co-Fe \cite{Schoen16}, while in metallic materials, the coupling to electrons is more relevant\cite{Kormann14}. 
Fig.\,\ref{fig3} illustrates a theoretically computed DOS for  YIG \cite{Wang20} with a fit that contains eighteen Lorentzians. (Parameters are displayed in Table~2 in the SM.) In this fit, a few negative amplitudes $A_j$ in Eq.\,\eqref{eq:Dw-Lorentzian-sum} are needed to reproduce the gap near $16\un{THz}$, however, the total $D_{\omega}$ remains positive.
Using additional information of the typical Gilbert damping parameter for this material \cite{Krysztofik17}, also the peak amplitude $A_1$ can be determined (see the SM).

More generally, via Eq.~\eqref{eq:scalar-Dw-Cw} the parameters of the multi-peak DOS \eqref{eq:Dw-Lorentzian-sum} immediately specify the functional form of the coupling $C_{\omega}$ of a system to a phononic bath in real materials.
This second result of the paper will be useful for modelling the Brownian motion of spins \cite{Anders20, Coffey20} and in applications such as quantum information processing with solid-state spin systems \cite{Hegde20}. 

\section{Conclusion} 

We have derived the general relation \,\eqref{eq:Dw-Cw}
that translates the function $\cal C_{\omega}$, determining the coupling of a generic system to a bosonic bath at various frequencies, into the density of states $D_{\omega}$ of the latter. This was achieved by evaluating the memory kernel of dynamical bath variables in two equivalent approaches.
Several applications of the relation were then discussed.
We demonstrated how for systems damped by phonons in $3$D with a frequency-independent $g$, Debye's quadratic DOS captures the same physics as a linear coupling function $C_{\omega}$ which corresponds to an Ohmic spectral density. 
Secondly, we have established how to infer $C_{\omega}$ from the measured DOS of a material, such that it reflects the specific properties of the material. 
Given that real materials have densities of states with multiple peaks, the typical picture which emerges from our general relation \eqref{eq:Dw-Cw} is that the coupling function is non-Ohmic and memory effects in the system dynamics become important.
The corresponding time scales (in the ps range, e.g., for gold in Fig.\,\ref{fig2}\,(b)) can be conveniently determined by fitting multiple Lorentzians to the bath DOS.

Future work could address how to extend relation \eqref{eq:Dw-Cw} to systems interacting with multiple independent baths. This should be suitable for non-equilibrium settings involving different temperatures \cite{Millen2014}, as used in heat transport \cite{Dhar08}. The impact of memory may also change the behaviour of systems like superconducting qubits or two-level systems that are in contact with two baths \cite{Senior20, Segal05}. 

\section{Acknowledgments}
We thank Jorge A. Mu\~{n}oz, Lisa M. Mauger and Brent Fultz for sharing their experimental data. We would also like to thank Joseph Barker, Luis Correa, and Simon Horsley for illuminating discussions, and Matias Bargheer for comments on an early draft of the paper. We gratefully acknowledge funding for this research from the University of Potsdam.

\clearpage 
\begin{figure*}[tp]
 \includegraphics[width=\textwidth]{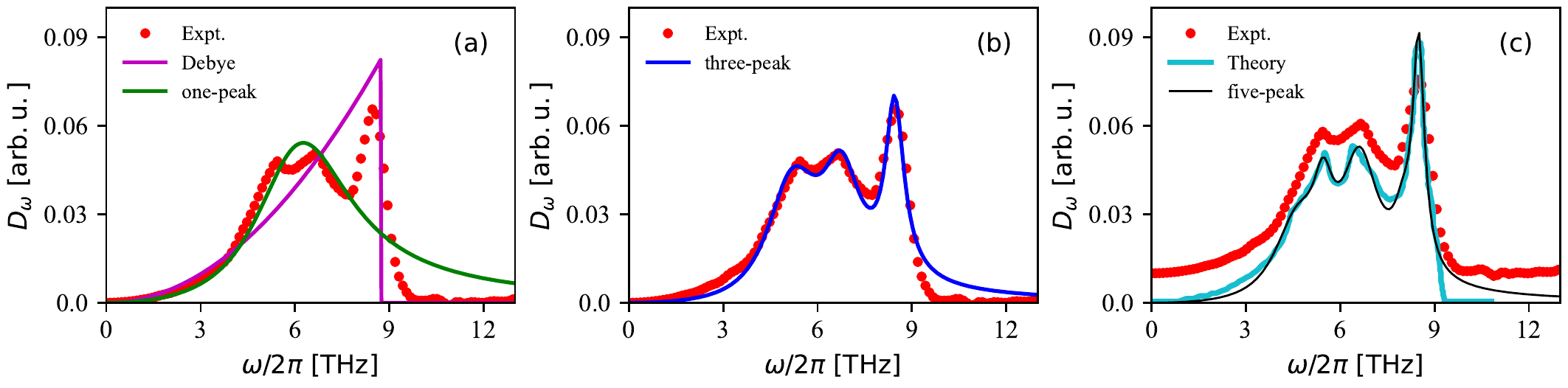}\vspace*{-05mm}
 \caption[]{Various fits for the phonon DOS in iron, compared to experimental measurements (red dots in (a), (b) and (c)) reported in Ref.\cite{Mauger14}. (a) The pink and green curves arise from Debye model(Eq.\,(19)) and the single-peak Lorentzian ($\nu = 1$ in Eq.\,(22)), respectively. (b) A three-peak Lorentzian  (Eq.\,(22) fits the experimental data well. (c) The black curve is a five-peak Lorentzian Eq.\,(22) fitted to the theoretical phonon DOS (cyan curve) reported in Ref.\cite{Mauger15}. Experimental data are shifted vertically for clarity. The fit parameters for the Lorentzian peaks are given in Table~\ref{tab:fit-to-Fe}.\vspace*{-05mm}
 }
\label{fig4}
\end{figure*}

\section{Supplementary Material}
Fig.\,\ref{fig4} shows the phonon DOS for iron, a well-known magnetic conductor.
In this figure, we use experimental data reported in Ref.\cite{Mauger14} and microscopic theoretical results calculated in Ref.\cite{Mauger15}. 
Fig.\,\ref{fig4} (a) shows fits with the Debye DOS [Eq.\,(19) of the main paper] and a Lorentzian DOS [Eq.\,(22) with a single peak]. The Debye cut-off frequency is taken from the Debye temperature ($420 \,\un{K}$) according to $\hbar\omega_D = k_B T_D$.
Both models reproduce the quadratic scaling at low frequencies, but the Lorentzian also captures the first two peaks. It does not properly capture the data for large frequencies, however. 

In Fig.\,\ref{fig4} (b), a three-peak Lorentzian  [Eq.\,(22)] fits the measured DOS spectrum remarkably well. It slightly deviates at higher frequencies because the Lorentzians decay less rapidly. Therefore, one can conclude that all three models are reliable fits to the measured data at low frequencies. However, if details are required in a broader frequency range (i.e. at shorter time scales), the three-peak Lorentzian DOS will be the better choice. 

Fig.\,\ref{fig4}(c) illustrates the additional structure in the DOS that becomes visible when fitting to the theoretical DOS of Ref.\cite{Mauger15}.
Here, we fitted a five-peak Lorentzian [see Table~\ref{tab:fit-to-Fe}] to reproduce two additional shoulders. Compared to the experimental data, these are probably hidden by the finite resolution of the measurement. Taking experimental data with such a broadening at face value, one would therefore produce a fitted coupling function with a somewhat shorter memory time compared to the real material. Fit parameters are given in Table~\ref{tab:fit-to-Fe}.

In Table~\ref{tab:fit-to-YIG}, we give the fitting parameters for the DOS of the material Yttrium Iron garnet (YIG). This is a magnetic material which has phonon-damped spin dynamics. Here, an absolute magnitude for the peak amplitudes can be extracted using the known Gilbert damping parameter $\eta \simeq 5 \times 10^{-4}$ \cite{Krysztofik17} and the electron gyromagnetic ratio $\gamma_e \simeq 28\times 10^{9} \un{Hz/T}$ \cite{Hauser16}. We consider for the system operator $\hat{\bf S}$ in Eq.\,(2), as in Ref.\cite{Anders20}, a spin vector ${\bf s}$ multiplied by the gyromagnetic ratio $\gamma_e$ (with $|{\bf s}| = \hbar/2$).
For the eighteen-peak fit, in Fig.\,3 of the main paper, we determine the absolute peak height to be $\Al_1  \simeq 71.14 \,\, \un{(rad\, THz\, T)^2/meV}$. The other fit parameters are given in Table~\ref{tab:fit-to-YIG}.
For the single-peak fit, which is easier to use in simulations, the absolute peak height is $\Al_1  \simeq 1194.19 \,\, \un{(rad\, THz\, T)^2/meV}$ and the other fit parameters are given in Table\,\ref{tab:fit-to-YIG}.

\begin{table}[tp]
    \vspace{-4ex}
    {\scriptsize
    \caption{Parameters of Lorentzian fits matched to experimentally measured Ref.\cite{Mauger14}  and theoretically calculated Ref.\cite{Mauger15} DOS for iron.}
        \label{tab:fit-to-Fe}
    \vspace{-1mm}
         \begin{center}
    \begin{tabular}{c|ccc}
    \multicolumn{4}{c}{single peak [Fig.\,\ref{fig4}(a)]}
    \\
            \hline
        \text{peak}& \text{frequency} & \text{width} & \text{ratio}\\
        $j$ & $\omega_{0,j}/2\pi\ [\un{{}THz}]$ & $\Gamma_{j}/2\pi\ [\un{{}THz}]$ & $\Al_{j}/\Al_1$
    \\
    \hline
        1 & 6.27 & 3.71 & 1.00 \\
    \hline
    \multicolumn{4}{c}{}\\
    \multicolumn{4}{c}{three peaks [Fig.\,\ref{fig4}(b)]}
    \\        
    \hline
        1 & 5.23 & 2.04 & 1.00 \\
        2 & 6.77 & 1.74 & 0.50 \\
        3 & 8.45 & 0.71 & 0.62 \\
        \hline
    \multicolumn{4}{c}{}\\
    \multicolumn{4}{c}{five peaks [Fig.\,\ref{fig4}(c)]}\\
        \hline
        1 &4.67 & 1.87 & 1.00 \\
        2 & 5.46 & 0.74 & 0.34 \\
        3 & 6.63 & 1.41 & 1.20 \\
        4 & 8.03 & 0.78 & 0.27 \\
        5 & 8.49 & 0.44 & 0.68 \\
        \hline
    \end{tabular}
    \end{center}
    \vspace{-05mm}
    }
\end{table}
\vspace*{-10mm}
\begin{table}[bp!]
    \vspace{-7ex}
        {\scriptsize
    \caption{Fit parameters of single-peak and eighteen-peak Lorentzians -- shown in  Fig.\,3 of the paper -- matched to the theoretical DOS for YIG reported in Ref.\cite{Wang20}.}
    \label{tab:fit-to-YIG}
    \vspace{-1mm}
    \begin{center}
    \begin{tabular}{r|rrr}
        \multicolumn{4}{c}{single peak }
    \\
            \hline
        \multicolumn{1}{c|}{\text{peak}}& \multicolumn{1}{c}{\text{frequency}} & \multicolumn{1}{c}{\text{width}} & \multicolumn{1}{c}{\text{ratio}}\\
        $j$ & $\omega_{0,j}/2\pi\ [\un{{}THz}]$ & $\Gamma_{j}/2\pi\ [\un{{}THz}]$ & $\Al_{j}/\Al_1$
    \\
    \hline
        1 & 5.91 & 12.4 & 1.00 \\
    \hline
    \multicolumn{4}{c}{}\\
    \multicolumn{4}{c}{eighteen peaks}
    \\        
        \hline
        1 & 2.56 & 0.99 & 1.00 \\
        2 & 3.66 & 1.35 & 16.20 \\
        3 & 4.89 & 1.22 & 10.10 \\
        4 & 6.45 & 0.55 & 1.47 \\
        5 & 7.16 & 0.99 & 7.75 \\
        6 & 8.10 & 1.20 & 10.60 \\
        7 & 9.20 & 1.18 & 11.50 \\
        8 & 10.20 & 0.54 & 1.70 \\
        9 & 10.80 & 1.82 & 11.30 \\
        10 & 12.60 & 1.67 & 33.20 \\
        11 & 13.70 & 0.83 & 9.07 \\
        12 & 13.80 & 3.80 & --86.60 \\
        13 & 14.40 & 1.30 & 19.60 \\
        14 & 16.10 & 1.07 & --13.70 \\
        15 & 16.40 & 1.83 & 40.10 \\
        16 & 18.70 & 1.46 & 6.08 \\
        17 & 20.10 & 0.94 & 4.27 \\
        18 & 20.90 & 0.45 & 2.20 \\
        \hline
    \end{tabular}
    \end{center}
}
\end{table}


\end{document}